# Controlling Nanogap Quantum Dot Photoconductivity

# through Optoelectronic Trap Manipulation

*Lauren J. Willis[†], Jessamyn A. Fairfield[†], Tali Dadosh[†], Michael D. Fischbein, Marija Drndic\**

Department of Physics and Astronomy, University of Pennsylvania, Philadelphia, Pennsylvania, 19104.

\*: To whom correspondence should be addressed: e-mail, drndic@physics.upenn.edu; phone, 215-898-5810; fax, 215-898-2010.

[†]: These authors contributed equally to this work.

ABSTRACT

Nanoscale devices are being extensively studied for their tunable electronic and optical properties, but the influence of impurities and defects is amplified at these length scales and can lead to poorly understood variations in characteristics of semiconducting materials. By performing a large ensemble of photoconductivity measurements in nanogaps bridged by core-shell CdSe/ZnS semiconductor nanocrystals, we discover optoelectronic methods for affecting solid-state charge trap populations.  We introduce a model that unifies previous work and transforms the problem of irreproducibility in nanocrystal electronic properties into a reproducible and robust photocurrent response due to trap state manipulation. Because traps dominate many physical processes, these findings may lead to improved performance and device tunability for various nanoscale applications through the control and optimization of impurities and defects.

One of the grand challenges of nanoscale systems is to control local fluctuations and disorder. As size decreases, the importance of individual defects and impurities grows, and they can cause unpredictable and undesired changes in behavior. Charge traps are ubiquitous and can affect a variety of systems: electronic states in graphene,[1] charge transport in carbon nanotubes,[2] photoluminescence intermittency in semiconducting nanocrystals and nanorods,[3,4] and telegraph noise in resistance.[5] Local fluctuations often act as 'hidden variables' that foil attempts at quantitative property measurement and interpretation. Hence, the discovery of ways to detect and control trap behavior will hasten progress in the field of nanoscience.

Nanocrystal quantum dots are being studied for a broad range of optoelectronic applications, including the realization of tunable and efficient photodetectors,[6-10] solar cells,[11,12] and light emitters.[13] CdSe/ZnS core-shell nanocrystals are particularly interesting because of their high quantum yield, large bandgap tunability across the visible spectrum, and well-established synthesis protocols.[14-16] Quantum dot arrays are light-sensitive artificial solids that serve as flexible model systems for the study of basic transport phenomena, arising from the interplay between the recombination-rate limited and thermally-activated charge transport mechanisms.[10,17-19] Prior studies have considered the presence of traps in these systems, but were not able to quantify or control them.

Here we demonstrate a robust and reproducible procedure for controlling the trap population in nanocrystal nanogap devices and show that qualitatively different photocurrent behaviors can be produced depending on how traps are initialized prior to a measurement. Electric field induced population and optically induced depopulation of traps can reverse the temperature dependence of the photoconductivity. We present a model that explains the role of traps and the importance of measurement sequence. Our method for dynamically controlling trap populations achieves optimized photodetector sensitivity at low or high temperatures for light sources, photovoltaics, electronics, and other applications. Moreover, we show that a range of temperature-dependent behaviors previously attributed to material differences are reproducible in a single device and provide a possible explanation for contradictory reports of the temperature dependence of photoconductivity in the literature.[10,17,19-22] These results carry implications for past and future experiments and may inspire analogous procedures for trap manipulation in other systems.

Figure 1a shows the schematic representation of the photodetector device, based on nanocrystals in a nanogap electrode geometry. Electrodes made of 3 nm of nickel and 30 nm of gold, separated by only 20-30 nm, ~4 nanocrystal diameters, are patterned using electron beam lithography on a silicon nitride ($Si_3N_4$) membrane.[23] The membrane is compatible with high-resolution structural characterization using transmission electron microscopy (TEM), which allows us to confirm that the nanogaps did not have any metallic debris, as shown in Figure 1b. An optical image of lithographic features on a typical device is shown in Figure 1c. For more details on device fabrication, please see Supporting Information Section 1. One advantage of nanoscale gaps is that the application of relatively small voltages yields high electric fields in the gap area. For a 20 nm gap with a bias voltage of 2V, the field strength experienced in the 2000 $nm^2$ of active area is $10^8$ V/m. The active area of these photodetectors in comparison to previous literature[17-19] is decreased by six orders of magnitude in area and decreased in gap size by two orders of magnitude.

We used Sigma-Aldrich CdSe/ZnS core-shell nanocrystals, as shown in Figure 1d, which had an average diameter of 5.2 nm, and a shell thickness of ~0.2 nm or ~1 monolayer. Before any treatment, the nanocrystals have a primary absorption peak at 610 nm and emit at 640 nm. The nanocrystals were capped with a mixture of hexadecylamine and trioctylphosphine ligands to prevent aggregation and passivate surface traps. Five microliters of the nanocrystal solution were dropcast onto the chip and allowed to dry, forming a multilayer nanocrystal film on the surface. For more characterization of the nanocrystals used, see Supporting Figures S1-S3.

Electrical measurements were performed in either a modified Janis VPF-700 or ST-100H cryostat operated at ~5×10$^{-7}$ torr. Nanogaps were wire bonded to a ceramic chip carrier thermally coupled to a copper cold finger and fitted into a Macor socket electrically addressed by silver-soldered wires making it compatible with high temperature operation, which coupled the source and drain pins to two independent BNC breakout boxes. Voltages were applied with a Yokogawa 7651 programmable DC source; current signals were amplified and filtered by a Keithley 428 current amplifier and measured with an Agilent 34401A digital multimeter. For all measurements at fixed laser intensity, the I-V characteristics for each device were measured by sweeping the voltage across the nanogap from 0V to 2V to -2V and back to 0V, with a typical cycle taking 200 seconds. This gives a maximum voltage drop per nanocrystal of 0.5 V. Prior to nanocrystal deposition and after TEM inspection, the bare devices were cleaned with an $O_2$ plasma, then the conductance and photoresponse of the bare devices were tested. The dark current of devices was measured by performing an I-V sweep with the nanogap in the dark, and photocurrent was measured by performing an I-V sweep while the nanogap was illuminated by a continuous wave 532 nm diode laser. In each measurement set, we measured at room temperature the dark current of all devices on a chip, and then the photocurrent of the same devices; the device was then cooled with either liquid nitrogen or liquid helium, and both dark current and photocurrent were measured again at low temperature. Changing measurement order, e.g. performing low-temperature measurements first and room-temperature measurements second, did not affect current characteristics.

The samples were annealed *in situ* because annealing has been shown to increase photocurrent in nanocrystal solids[17-19,22] by reducing interparticle separation and lowering tunneling barriers. We detected photocurrent in 17 nanogaps, and no dark current signal above the maximum noise floor of ~0.03 pA at 295K and ~0.15 pA at 78K in 70% of devices after annealing up to 573K. From TEM imaging we confirmed that nanogaps did not have any metallic debris that could contribute to the dark current, and this was consistent with our subsequent I-V characterization of the nanogaps. More importantly, all of our photocurrent is primary, which means it is a result of direct exciton generation in the nanocrystals and that there is no measurable charge injection from the metal electrodes into the nanocrystal film.[21] For more details about the effect of annealing on photocurrent and dark current measurements, see Supporting Information Section 3 and Figures S4-S6.

Figure 2 shows the I-V response under 532 nm illumination at different temperatures of two different nanogaps. The photocurrent is well described by a second-order polynomial in voltage. Measurements at other wavelengths show consistent behavior once temperature-dependent absorption shifts are accounted for; see Supporting Figure S7 for I-V characteristics obtained using different illumination wavelengths. Over all measured devices, the initial room temperature photocurrent was in the range of 0.1 to 50 pA, with a mean value of ~5 pA, and the initial low temperature photocurrent was in the range of 0.1 to 240 pA, with a mean value of ~30 pA. A histogram of photocurrent values is shown in Supporting Figure S8. The large variation in measured photocurrent is probably due to the small number of nanocrystals in the nanogap; thus the variations in each individual nanocrystal are not averaged out. Additionally, film thickness within the nanogap and the energy barrier at the contacts between the nanocrystals and the electrodes may vary. Transport through the nanocrystals inside the gap dominates the photocurrent, while nanocrystals outside the gap region have a negligible contribution.[6] Although more than just the gap area is illuminated, outside the gap, the high barrier to interparticle transport and the low field prevent significant contribution to photocurrent. See Supporting Figures S9 and S10 for optical and TEM images of measured nanogaps.

We observed that the magnitude of the measured photocurrent depends on the recent illumination history of the device. Even more strikingly, some nanogaps showed photocurrent that was higher at 295K than at lower temperatures, while other nanogaps on the same chip and under equivalent conditions showed the opposite. Moreover, if the nanocrystals were illuminated overnight and voltage was applied (hereafter referred to as a *laser voltage treatment*), the low-temperature photocurrent was enhanced, whereas if the nanocrystals were left in darkness overnight and voltage was applied (hereafter referred to as a *dark voltage treatment*), the low-temperature photocurrent was suppressed. If voltage was not applied while the sample was left in darkness overnight, the photocurrent magnitude returned to its initial value. This trend was repeatable over a measurement period of several months. The photocurrent at 295K followed the same trend as the low-temperature photocurrent in ~75% of devices, but the effect was smaller (~10-30% of the photocurrent change at 78K).

To best quantify the photocurrent increase or decrease with temperature, T, it is convenient to define the *relative photocurrent ratio* $R = I_{78K}/I_{295K}$, of the low-temperature photocurrent, $I_{78K}$, and the room-temperature photocurrent, $I_{295K}$. This is analogous to the *relative resistance ratio* between the low- and room-temperature resistance in metals, commonly used as a criterion of metal purity; if the photoconduction in nanogaps were Ohmic, resistance would be well defined, and then $R$ would be the same as that defined for metals. Each ratio $R$ was calculated for one cool-down cycle of measurements taken in a single day. The relative photocurrent ratio has two distinguishable regimes: if $R < 1$, this means that the photocurrent increased with T, and if $R > 1$, the photocurrent decreased with T. In the discussion below, $I_{78K}$ and $I_{295K}$ were calculated as averages of photocurrent magnitudes for the maximum electric field applied across the nanogaps, corresponding to voltages -2V and 2V. The following conclusions hold qualitatively for other voltages, and also apply independently of annealing temperature. Examples of nanogaps with $R = I_{78K}/I_{295K}$ smaller or greater than 1 are shown in Figure 2a ($R = 2.2$) and Figure 2b ($R = 0.1$). Out of the seventeen nanogaps, fifteen initially showed $R > 1$ and two showed $R < 1$. A histogram of $R$ values for 532 nm illumination and a comparison of $R$ values for both 532 and 650 nm illumination are given in Supporting Figures S11 and S12. As measurements progressed, illumination history was observed to affect this ratio, so that $R$ could be switched from less than 1 to greater than 1 or vice versa in a single nanogap. A sample table of the change in relative photocurrent ratios after laser and dark voltage treatments is given in Supporting Table S1.

Figure 3 shows a histogram of the *change* in R from $R_{initial}$ to $R_{final}$ from 70 measurements over all devices after laser and dark voltage treatments. We use a logarithm transformation to write the change in R on the x-axis as $\ln\left(\dfrac{R_{final}}{R_{initial}}\right)$. This manner of representing the change in R is informative because $\ln\left(\dfrac{R_{final}}{R_{initial}}\right) = -\ln\left(\dfrac{R_{initial}}{R_{final}}\right)$, meaning that an increase or decrease of $R$ by the same factor is represented on this scale symmetrically around zero; $\ln\left(\dfrac{R_{final}}{R_{initial}}\right) = 0$ means that $R$ does not change. There are two distinct distributions in this histogram, showing that device behavior after a laser and dark voltage treatment is clearly separated. The laser voltage treatment increases the ratio by an average factor of 2.2, meaning that $R_{final} > R_{initial}$. The dark voltage treatment decreases the ratio by a factor of 10, meaning that $R_{final} < R_{initial}$. We have also observed this effect in a large gap with an active area of ~109 μm$^2$ (~43.6 x 2.5 μm), implying that this effect is independent of device size.

This demonstrated change in ratio $R$ with laser or dark voltage treatments shows that the temperature dependence of conductivity is controlled by the measurement protocol. Consequently, all such measurements on nanocrystal arrays must be framed in the context of the sample measurement history in order to be properly interpreted. This consideration may explain apparent discrepancies in the reported temperature dependence of observed photocurrent.[10,17,19-22] As we will demonstrate, localized charge carriers in the nanogap can measurably affect the temperature dependence of photoconductivity. The manipulation of trap states by optically stimulated emptying or voltage induced populating can then be used to control device performance.

To understand the underlying mechanism, we must first look at the energy levels through which the charge carriers travel. Figure 4a shows energy levels for the electrodes with a single nanocrystal between them; the shortest charge carrier path in our devices includes four nanocrystals. The carrier tunneling between nanocrystals can be lost by recombining with other oppositely charged carriers through radiative or nonradiative recombination, which usually corresponds to free recombination or recombining with trapped carriers at recombination centers, respectively. The primary photocurrent in a semiconductor is given by

$$I(E,T) = eFG \,, \tag{1}$$

where $e$ is the charge of an electron, $F$ is the exciton generation rate, and $G$ is the number of free charge carriers created that pass between the electrodes for each photon absorbed, which is also called the photoconductive gain.[18,19,24] $F$ is defined by

$$F = \Phi a \eta(E,T),^{19} \tag{2}$$

where $\Phi$ is the excitation flux, $a$ is the film absorption, and $\eta(E,T)$ is the field-dependent exciton separation efficiency. $\eta(E,T)$ is defined in terms of the relevant rates that affect exciton recombination and transport:

$$\eta(E,T) = \frac{k_E(E,T)}{k_E(E,T) + k_R(T) + k_N(T)}, \tag{3}$$

where $k_E(E,T)$ is the field-dependent rate of charge carriers escaping to neighboring nanocrystals or electrodes, $k_R(T)$ is the rate of charge carriers radiatively recombining, and $k_N(T)$ is the rate of charge carriers nonradiatively recombining.[24]

The tunable temperature dependence of the observed photocurrent can be explained by the relative magnitudes of the rates $k_R(T)$, $k_N(T)$, and $k_E(E,T)$ involved, shown in Figure 4a, and their temperature dependence. $k_R(T)$ increases with increasing temperature because thermal energy magnifies the probability of free charge carriers to recombine, causing photocurrent to decrease with increasing temperature.[25] $k_N(T)$ decreases with increasing temperature, since at high temperature, less nonradiative recombination occurs because charge carriers easily escape from traps with thermal energy, which causes photocurrent to increase with increasing temperature.[24] The contribution from both radiative recombination and the number of traps is constant over these measurements and fixed for a given sample, but the contribution from trap states depends on trap population, which can be adjusted by laser and dark voltage treatments.

Before any treatment, the system has a number of occupied trap states that is defined as the steady state, as in Figure 4a. The laser voltage treatment creates many charge carriers that recombine with carriers in trap states, causing traps that are occupied in steady state to be emptied, as in Figure 4b; this effect of optically stimulated trap emptying in our nanogaps is similar to an analogous phenomenon well documented in the semiconductor literature.[24] The laser voltage treatment eliminates many charge carriers, even in energetically favorable traps, and fewer charge carriers recombine with trapped charges. This increases photocurrent temporarily, but over a few days of waiting time, the trap occupancy returns to its steady state value, as energetically favorable traps are repopulated, causing photocurrent to return to a steady state value as well. Conversely, the dark voltage treatment repeatedly sweeps the voltage, increasing the population of charge carriers in the traps, as in Figure 4c. Thus, after dark voltage treatment even energetically unfavorable traps are populated; the populated traps capture more carriers and cause them to recombine, temporarily decreasing photocurrent. Over a few days of waiting time the trap occupancy returns to its steady state value, as charge carriers in some traps escape using thermal energy, causing photocurrent to return to a steady state value as well. To summarize, by applying the laser and dark voltage treatments, the trap population is modified, which allows tuning of the photocurrent response; this has a greater effect at low temperature because traps can be emptied using the larger thermal energy at room temperature. Relevant processes for photogenerated electrons in the conduction band are shown in detail in Figure 4. The photocurrent temperature dependence can be tuned using these effects, and the resultant adjustability is robust even when other variables are changed.

Measuring photocurrent dependence on laser intensity at a fixed wavelength also supports the trap-based model in explaining the adjustable photocurrent dependence on temperature. While initial photoconductivity measurements were taken with a fixed illumination intensity of ~65 mW/cm$^2$, intensity was later varied between 1.6 to 120 mW/cm$^2$ at both 78K and 295K. As illumination intensity was varied, the current was measured at a constant voltage of 1V, which corresponds to $3 \cdot 10^5$ V/cm. The laser and dark voltage treatments had little effect on the intensity dependence at 295K, but had a greater effect at 78K. The treatments can result in an inversion of the temperature dependence of the photocurrent for a wide range of intensities, see Supporting Figure S13.

At 295K, the intensity dependence of the photocurrent always followed a power law, $I_{photo} \propto$

*Intensity*$^{\alpha}$, where $\alpha = 0.82 \pm 0.02$ over seven measurements, as shown in Figure 5. This is consistent with previous room temperature measurements on large arrays of core-shell nanocrystals yielding the same $\alpha$ value.[19] Intensity dependence of the photocurrent at 78K gives $\alpha = 0.96 \pm 0.02$ in agreement with the linear response at low temperature reported in the literature.[20] The specific value of $\alpha$ helps reveal the type of carrier dynamics present.

In order to understand the power law fit, it is instructive to examine the rate equation for $n$, the density of free electrons,

$$\frac{dn}{dt} = F - C(n + n_{trap})n \text{ .[18,24]}$$

(4)

Here, $C$ is the probability of an electron to be captured, and $n_{trap}$ is the density of trapped electrons. $n + n_{trap}$ represents the density of holes in the system which can recombine with free electrons, assuming a neutral nanocrystal. For a steady state system, $\frac{dn}{dt} = 0$ and we can rewrite $F$:

$$F \propto (n_{trap} + n)n \text{ .}$$

(5)

Substituting equation (2) into equation (5), we can write

$$\Phi \propto \frac{(n_{trap} + n)n}{a\eta} \text{ .}$$

(6)

For $n_{trap} >> n$, $\Phi \propto n$, but for $n_{trap} << n$, $\Phi \propto n^2$. Since $n \propto I_{photo}$, this means that for $n_{trap} >> n$, $I_{photo} \propto \Phi$, but for $n_{trap} << n$, $I_{photo} \propto \Phi^{0.5}$. Thus, $\alpha = 1$ implies monomolecular (first-order) carrier dynamics, whereas $\alpha = 0.5$ implies bimolecular (second-order) carrier dynamics.[24] First-order kinetics contribute more when the material has many recombination centers, such as deep hole traps, and when the material has a lower free electron concentration than in the bulk, as is the case in nanocrystals where the presence of surface traps is likely.[20] Contributions of surface and deep traps, which are only partially passivated by the shell and ligands, can cause a deviation of the photocurrent dependence on intensity from the expected dependence in a bulk solid, where bimolecular recombination would dominate giving $\alpha = 0.5$. Our measured exponent $\alpha = 0.8$ at room temperature implies that we observe a combination of first and second order recombination dynamics. However, at low temperature, we measured an exponent $\alpha = 1$, implying that we observe first-order recombination dynamics. The variation in the fitting exponent sheds light on the difference of recombination center density at each temperature, which supports our proposed model shown in Figure 4 and encompasses reported intensity dependence.[19,20,24]

In conclusion, we have created nanogap devices with CdSe/ZnS core-shell nanocrystals in the gap region; after annealing, these devices can be operated as photodetectors with tunable photoconductivity. We investigated the temperature dependence of photocurrent and found that it depends on the illumination history of the device. Recent laser illumination causes optically induced trap emptying and higher low-temperature photocurrent, while recent voltage cycling in the dark causes electric field induced trap population and lower low-temperature photocurrent. This can resolve existing discrepancies in the literature, demonstrating the difficulty in defining temperature dependence of photoconductivity for semiconducting nanocrystal systems. Additional research in this area could include investigation of trap depopulation timescales, dynamic response, dependence on nanocrystal material or size, and the optimization of treatment parameters. We find our controllable photocurrent temperature dependence to be robust over multiple wavelengths and intensities of laser excitation and suggest a route towards achieving maximal photodetector response at different temperatures. This approach of tuning the photocurrent response via trap population can be useful for nanocrystal device applications, such as sensors, solar cells, and light emitters, as well as aiding in the study of carrier dynamics and energy levels in nanoscale materials.

ACKNOWLEDGMENTS. We thank Prof. Jay Kikkawa for very helpful discussions. We also thank Prof. Catherine Crouch and Dr. Chris Merchant for useful comments on the manuscript. This work was supported by the DARPA Young Investigator Award HR011-08-1-0031 and partially by ONR (YIP N000140410489) and NSF (NSF Career Award DMR-0449533, NSF MRSEC DMR05-20020 and NSF NSEC DMR-0425780). L. W. acknowledges funding from the NSF-IGERT program (Grant DGE-SUPPORTING INFORMATION. Supporting information includes details of device fabrication; nanocrystal characterization, which includes TEM images, a histogram of nanocrystal size, and absorption and emission spectra; additional annealing characterization; dark current I-V plots and Arrhenius fits; I-V curves for different wavelengths of illumination; histograms of photocurrent values at 2V under 532 nm illumination; optical and TEM images of a chip after nanocrystal deposition, annealing, and measurement; a histogram of $I_{78K}/I_{295K}$ under 532 nm illumination, a histogram of $I_{78K}/I_{295K}$ comparing the effects of using 532 and 650 nm illumination, and a table of changing relative photocurrent ratio; and photocurrent temperature inversion plots.

FIGURE CAPTIONS.

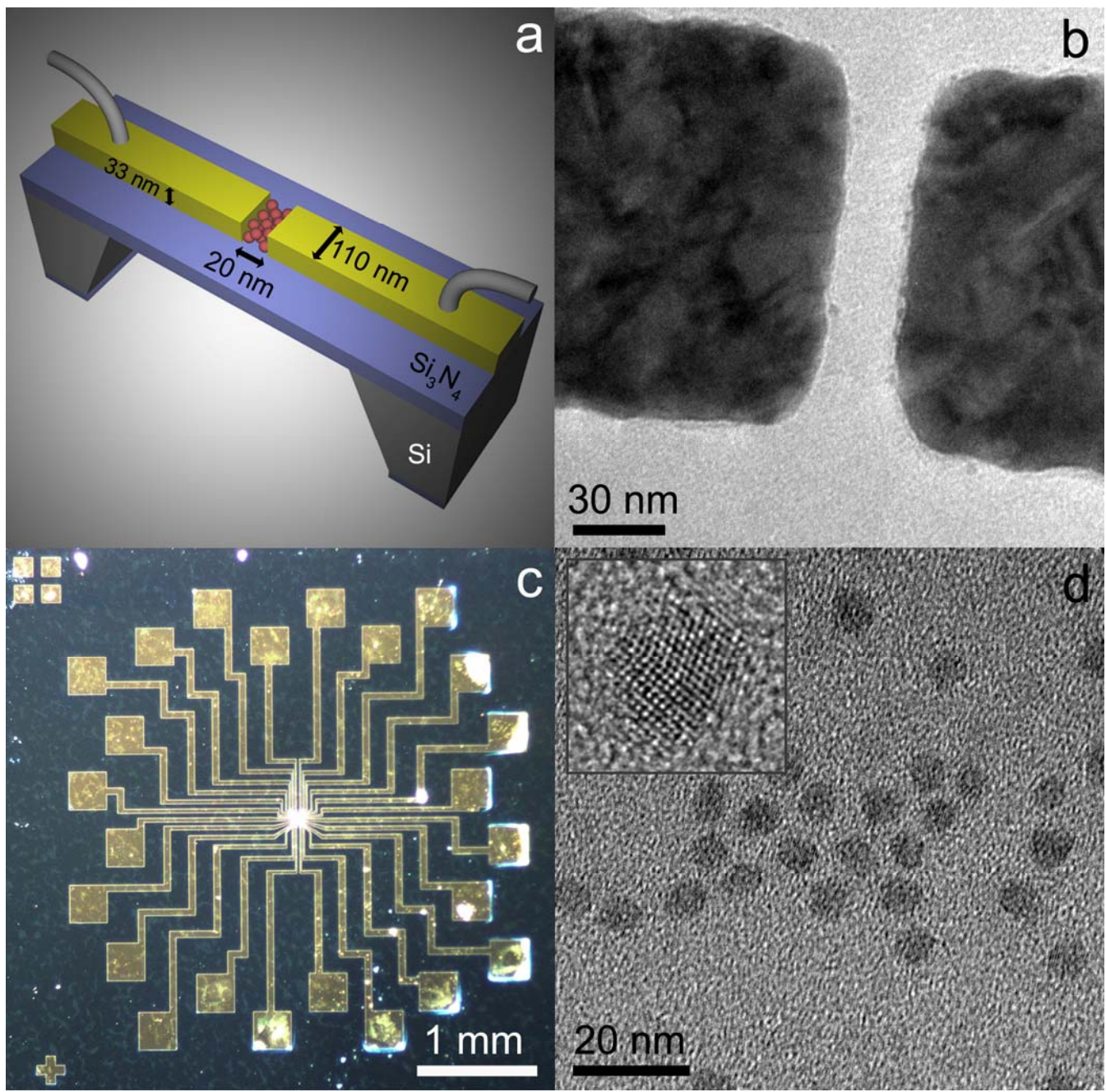

**Figure 1.** (a) Schematic of the photodetector nanogap device. Metal electrodes (3 nm of Ni and 30 nm of Au) separated by 20 to 30 nm are patterned on top of a 40 nm $Si_3N_4$ membrane that is supported by a Si wafer. Nanocrystals are deposited on the substrate and electrodes. (b) TEM image of the Ni/Au electrodes separated by 20 nm prior to nanocrystal deposition. (c) Optical image of the device with 12 electrode pairs. (d) TEM image of CdSe/ZnS nanocrystals with an average size of 5.2 ± 0.6 nm. Inset: Zoomed-in TEM image of a single nanocrystal.

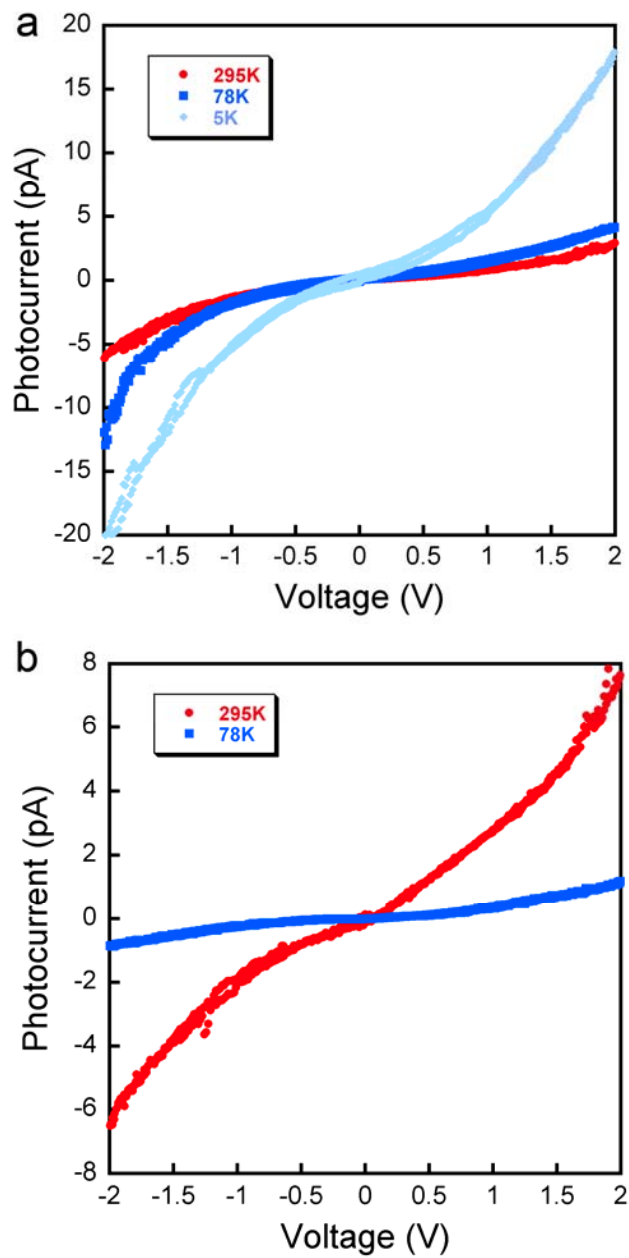

**Figure 2.** Photocurrent vs. voltage curves at 5K (light blue), 78K (dark blue), and 295K (red) for CdSe/ZnS nanocrystals in devices where the low temperature photocurrent is higher (a) or lower (b) than the room temperature photocurrent.

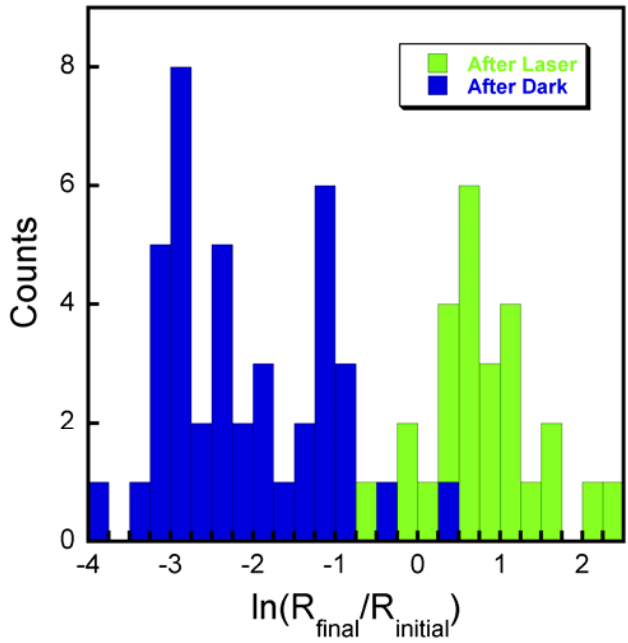

**Figure 3.** Histogram of the change in the final ratio with respect to the initial ratio, $R_{final}/R_{initial}$, on a logarithmic scale, with over 70 measurements including laser voltage treatments (green) and dark voltage treatments (blue). An increase or decrease of $R$ by the same factor is represented on this logarithmic scale symmetrically around zero. Two distinct distributions clearly show that $R$ increases after a laser voltage treatment and decreases after a dark voltage treatment on average 2.2 and 10 times, respectively.

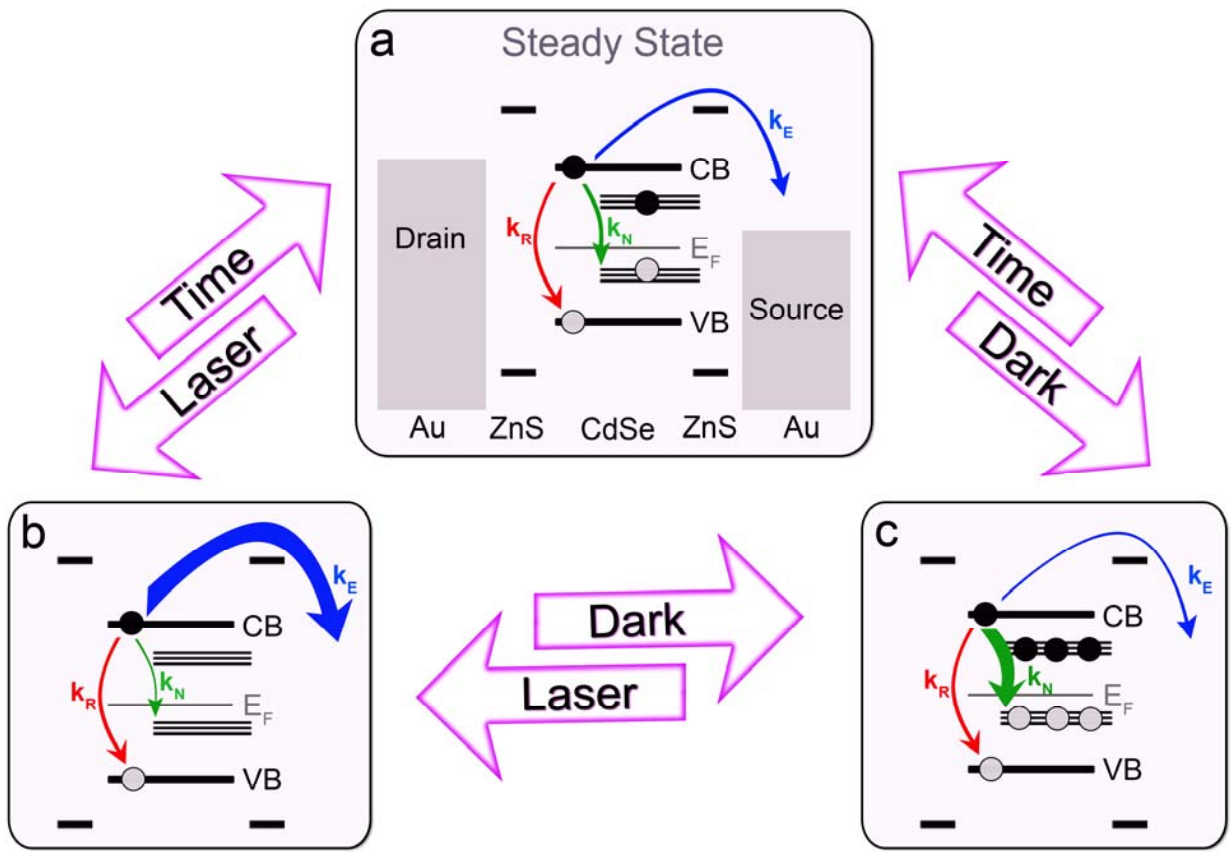

**Figure 4.** (a) The energy level diagram for a steady state CdSe/ZnS core-shell nanocrystal between two gold electrodes. Trap states above the Fermi energy, $E_F$, act as electron traps, whereas trap states below $E_F$ act as hole traps. An exciton is created by illumination of the sample, and it can either recombine radiatively with rate $k_R$, recombine nonradiatively via the trap states with rate $k_N$, or tunnel away from the nanocrystal at a rate $k_E$ related to the applied field $E$. (b) After applying laser voltage treatment, the traps are emptied which enhances $k_E$ and suppresses $k_N$. (c) After applying dark voltage treatment, the traps are filled, which suppresses $k_E$ and enhances $k_N$. Over a few days of waiting time, trap populations return to their steady state value, which also returns $k_E$ and $k_N$ to their steady state values. Hole processes are affected by the treatments in the same way, but are not shown.

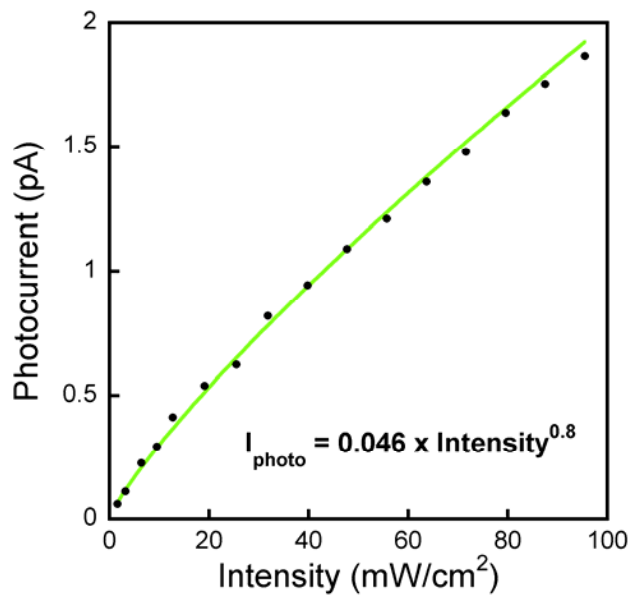

**Figure 5.** Photocurrent as a function of illumination intensity for 532 nm laser excitation measured at

$3 \cdot 10^7$ V/m at 295K. The green line is a power law fit to the data with an exponent of 0.8.

# Controlling Nanogap Quantum Dot Photoconductivity through Optoelectronic Trap Manipulation

Lauren J. Willis[†], Jessamyn A. Fairfield[†], Tali Dadosh[†], Michael D. Fischbein, Marija Drndic*

Department of Physics and Astronomy, University of Pennsylvania, Philadelphia, Pennsylvania, 19104.

*: To whom correspondence should be addressed: e-mail, drndic@physics.upenn.edu; phone, 215-898-5810; fax, 215-898-2010.

[†]: These authors contributed equally to this work.

## 1. Device fabrication

To fabricate the silicon nitride membranes, we use 500 μm thick <100> p-type silicon ($Si^+$) wafers with 100 nm of highly polished, low-stress amorphous silicon nitride ($Si_3N_4$) on both sides. These wafers were processed to produce many ~5 × 5 $mm^2$ chips, each with a 50 × 50 $μm^2$ region in its center where the $Si_3N_4$ membrane is freely suspended. The wafers are first coated on one side with a protective resist and on the other side with NR7 photoresist spun on at 3000 RPM for 42 seconds, and then baked at 115ºC for 3 minutes. The window mask is exposed to 365 nm light at 5 $mW/cm^2$ for 34 seconds and baked at 115ºC for 2 minutes. The wafer is developed in RD6 for 16 seconds, followed by a rinse with deionized water. To remove the silicon nitride, the wafer is then exposed to a $SF_6$ plasma etch in a Technics PeII-A Etcher at 50 W with a flow of 400 mtorr for 120 seconds. Finally, the wafer is exposed to a 1.5 M potassium hydroxide (KOH) wet etch at 130ºC. The KOH etches anisotropically through the silicon until the silicon nitride on the other side of the wafer is exposed, which takes approximately 18 hours. Once the etching is complete, the membrane window of 100 nm thick silicon nitride is further thinned to approximately 40 nm using another $SF_6$ plasma etch step.

Nanoelectrodes were patterned onto the window by electron beam lithography on an Elionix 7500-ELS, and 3 nm of nickel and 30 nm of gold were evaporated onto the devices. Nickel was chosen as an adhesive layer for the small features because it leaves almost no debris in nanoscale gaps. Large metal features and contact pads were added using optical lithography followed by thermal evaporation of 3 nm of Cr and 100 nm of Au. Chips are allowed to outgas overnight after lithography steps to avoid TEM contamination. Twelve nanogaps were patterned per chip, with the gap size measured with a JEOL 2010F field-emission TEM to be 20-30 nm.

## 2. Additional characterization of the CdSe/ZnS nanocrystals

Nanocrystal size was determined from high-resolution TEM images (Figure S1) as the average of 50 nanocrystals, to be 5.2 ± 0.6 nm (Figure S2). Each nanocrystal was measured twice, with the measurements of the same nanocrystal roughly perpendicular to each other. These 100 measurements were averaged, giving a mean value of 5.2 nm with a standard deviation of 0.6 nm. However, there may be a slight underestimation due to the increased difficulty in discerning the ZnS shell against the carbon grid background.

The absorption and emission spectra of nanocrystals in toluene solution were recorded using SpectraSuite from Ocean Optics (Figure S3). The excitation wavelength for the emission data was provided using an Ocean Optics LS-450 with a 470 nm LED and filter. The light source for the absorption spectrum was an Ocean Optics LS-1 with a tungsten halogen bulb. The main excitonic peak was at 610 nm and the emission peak was at 638 nm. The reported quantum yield from Sigma-Aldrich is 30%.

Using the formula fitted by Yu et al.,[1] an absorption peak of 610 nm would correspond to a diameter of ~5.06 nm for a CdSe core nanocrystal. It has been shown that adding a ZnS shell not only broadens, but also redshifts the absorption peak, and the size of the redshift depends on the size of the nanocrystal.[2] We estimate a redshift of ~5 nm, so that the absorption peak for our CdSe core should be 605 nm. From Yu et al., the core diameter of our CdSe nanocrystals is ~4.81 nm. It should be noted that had we not taken into account the redshift caused by the shell, we would have overestimated the core size by ~0.25 nm. Knowing that the core radius is 2.4 nm and the actual nanocrystal radius, determined by TEM, is 2.6 nm, we find that the shell is ~0.2 nm, which is approximately one monolayer.

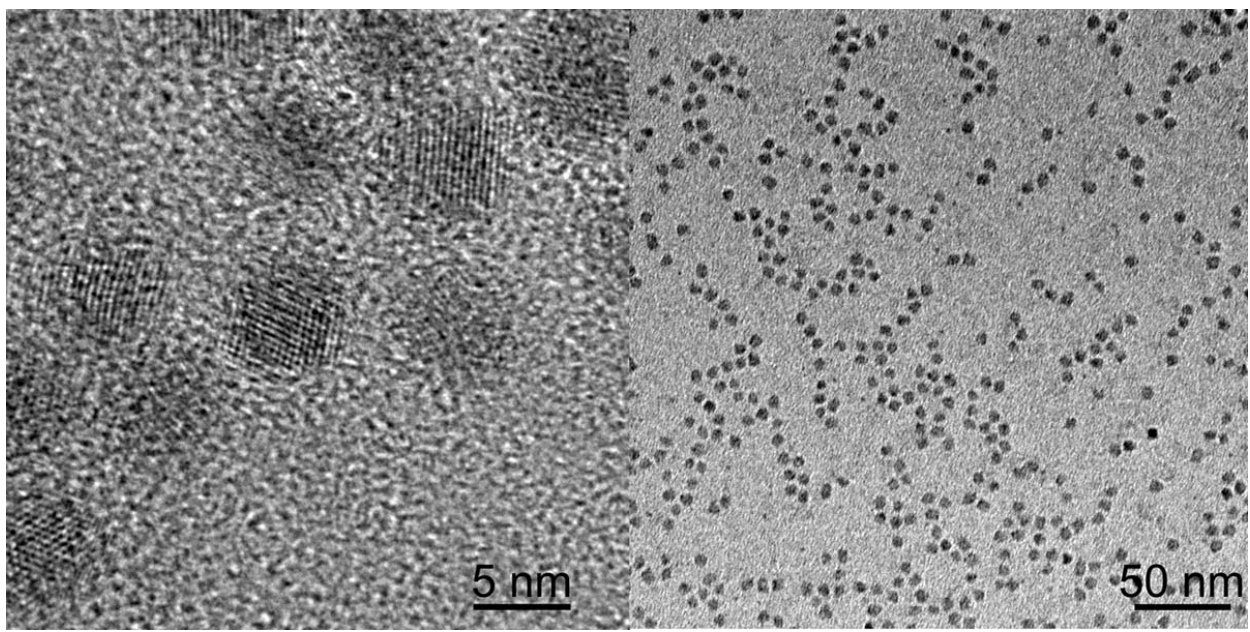

**Figure S1.** High-resolution TEM images of CdSe/ZnS nanocrystals on a carbon grid.

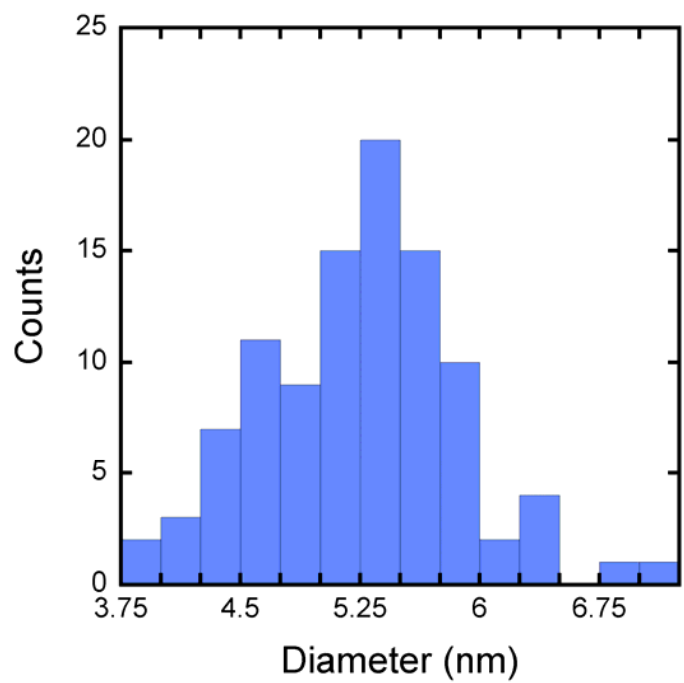

**Figure S2.** Histogram of 50 nanoparticles, each measured twice from TEM images.

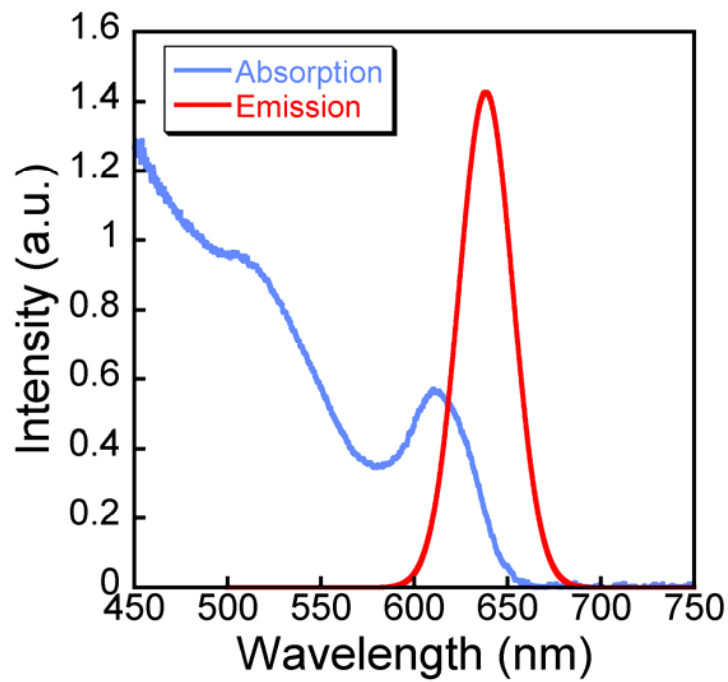

**Figure S3.** Absorption and emission intensity vs. wavelength curves for nanocrystals studied.

## 3. Effect of annealing

The percentage of devices that had measurable photocurrent increased with annealing temperature $T_a$. For $T_a = 423K$ the yield was 44%, for $T_a = 498K$ the yield was 66%, and for $T_a = 573K$ all devices showed photocurrent response. In devices that had photocurrent above the noise floor (0.03 pA at 295K and 0.15 pA at 78K), increasing the annealing temperature from 423K to 498K increased the observed photocurrent by 20 times on average. Above 498K, there was no measurable change in photocurrent magnitude or response characteristics. All data discussed were taken from 17 active nanogaps out of 20 total that were annealed at either 498K or 573K.

## 4. Dark current on bare device and device with nanocrystals

Dark current was measured on all nanogaps prior to any nanocrystal deposition, prior to any photocurrent measurements, and over the course of several months as devices were thermally cycled many times. Dark current was below the noise floor of our setup for 70% of devices measured. In Figure S4, we show examples of I-V characteristics for a bare nanogap measured at room temperature and for nanogaps with nanocrystals that have been annealed and measured at 78K and 295K.

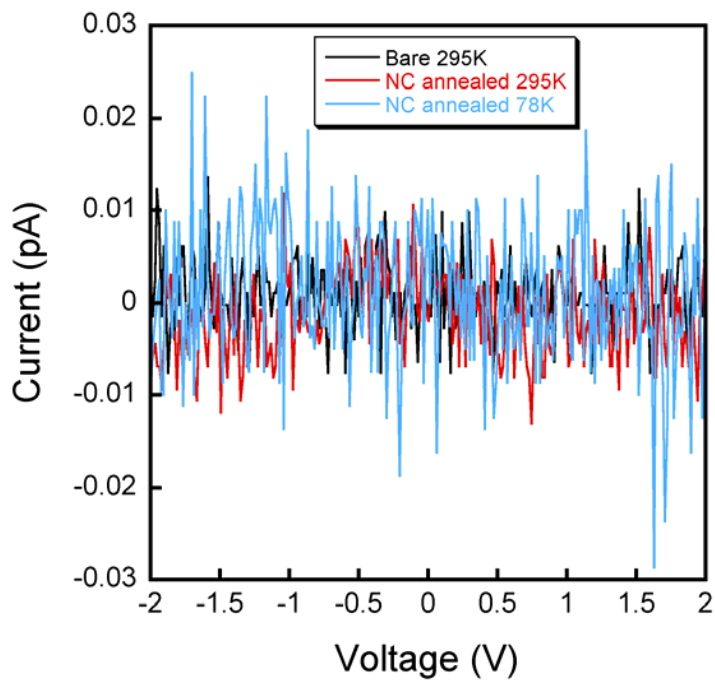

**Figure S4.** Representative I-V dark current curves for a bare device, and devices with nanocrystals annealed up to 573K and measured at 295K and 78K. More than 70% of devices showed no measurable dark current, making them primary photodetectors.

Less than ~30% of nanogaps, only those annealed at 573K, showed a very small dark current (~0.16 pA at 2V), 2-3 orders of magnitude smaller than the corresponding photocurrent. The dark current increases exponentially with voltage and can be empirically fitted to an exponential form $I_{dark} = \dfrac{V_o}{R_o} e^{\frac{V}{V_o}}$, where $R_o \sim 2 \times 10^{14}$ $\Omega$ and $V_o = 0.7$ $V$ are the characteristic resistance and voltage (Figure S5). This is in agreement with previously reported dark current measurements on micron-scale nanocrystal arrays.[3] Consequently, because there is no clear threshold of the dark current, this also implies that there is no intrinsic difference between primary and secondary photodetectors in these systems. Whether a photodetector is labeled "primary" or "secondary" is determined by the noise floor of the measurement setup.

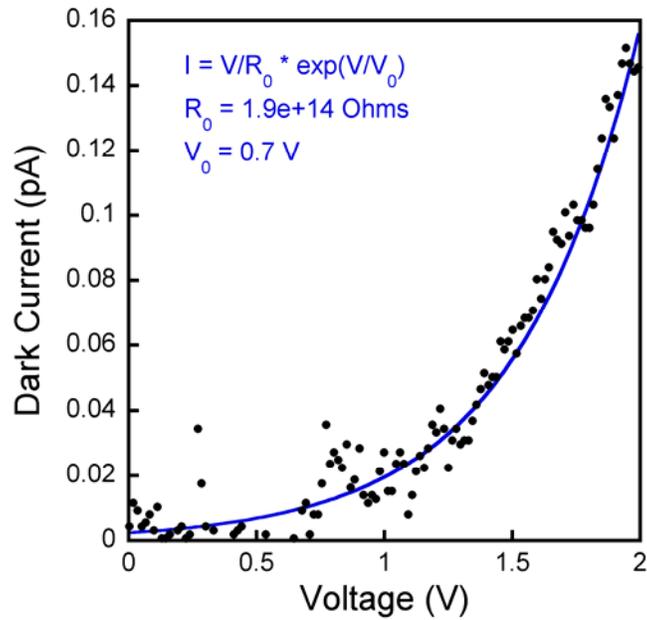

**Figure S5.** I-V curve for one of the few devices with measurable dark current.

Figure S6 shows the temperature dependence of the dark current, written as the zero-bias conductance $G$ vs. $1/T$, which is representative for the few devices that exhibited measurable dark current.

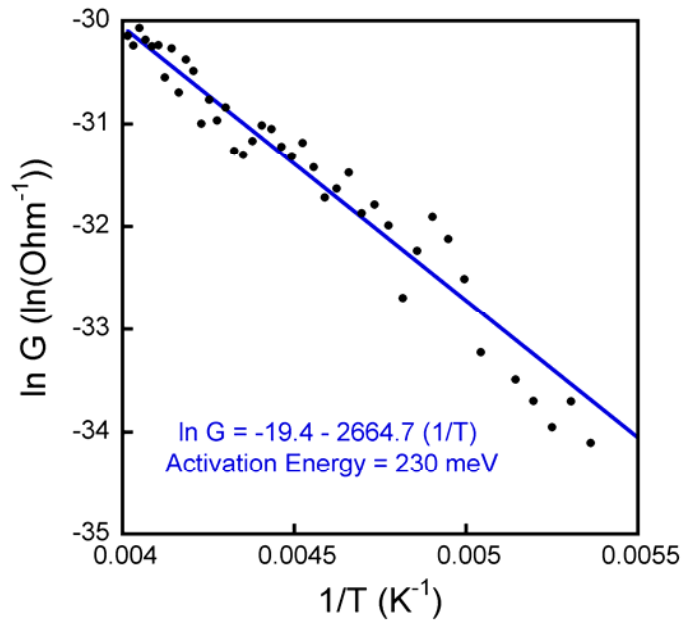

**Figure S6.** Sample Arrhenius plot with calculated activation energy.

The dark current is thermally activated and conductance was found to scale as $G \sim e^{-\frac{E_A}{k_B T}}$. We have measured a range of activation energies from ~70-230 meV, consistent with previously published results.[4]

# 5. Wavelength dependence of photocurrent

Examples of I-V sweeps at illumination wavelengths of 473 nm, 532 nm, and 650 nm and the positions of these laser excitation wavelengths on the absorption vs. wavelength curve are shown in Figure S7.

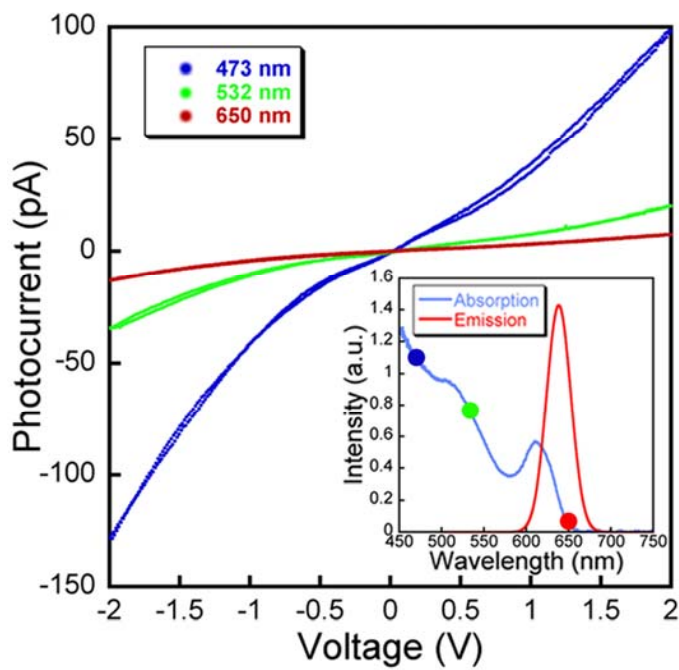

**Figure S7.** Representative photocurrent vs. voltage curves for 650 nm (red), 532 nm (green) and 473 nm (blue) laser excitations. Inset: Absorption and emission intensities vs. wavelength for CdSe/ZnS nanocrystals in solution. The blue, green, and red circles indicate the positions of the excitation wavelengths with respect to the absorption curve.

# 6. Statistics of photocurrent

Nanocrystal films at the nanoscale show some amount of nonuniformity because of the small number of nanocrystals in the gap area. Figure S8 shows the distribution of photocurrent magnitudes for all our measured data using a 532 nm laser.

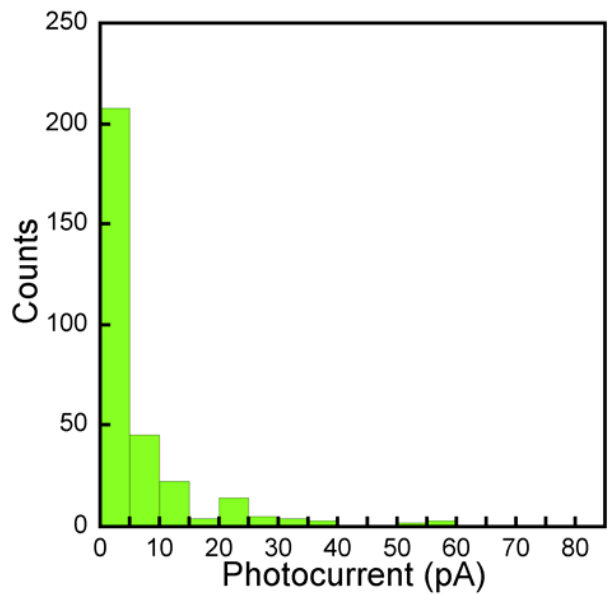

**Figure S8.** Distribution of photocurrent magnitudes for nanogap devices illuminated with 532 nm light. Histogram includes data from all treatments.

## 7. Representative images of a measured device

Figure S9 shows an optical image of an entire chip after nanocrystal deposition, annealing, and photocurrent measurements. Figure S10 shows two TEM images of different nanogaps after nanocrystal deposition, annealing, and photocurrent measurements. The nanocrystal film in the gap area appears blurred due to the thickness of the film.

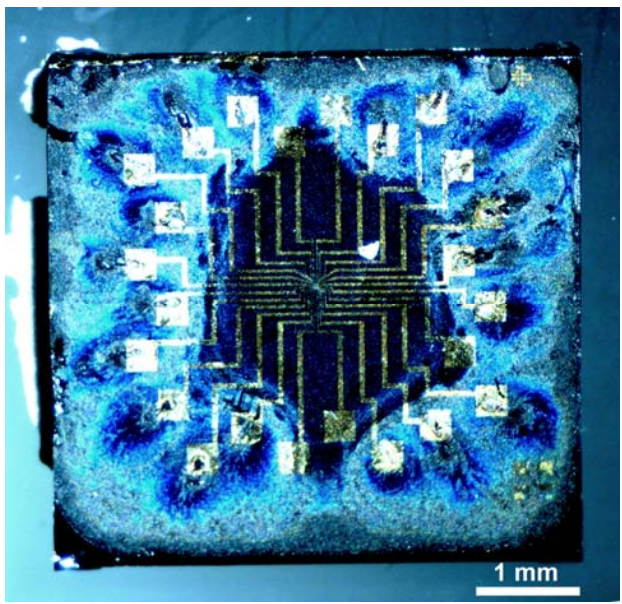

**Figure S9.** Optical image of a device with annealed nanocrystals after measurement.

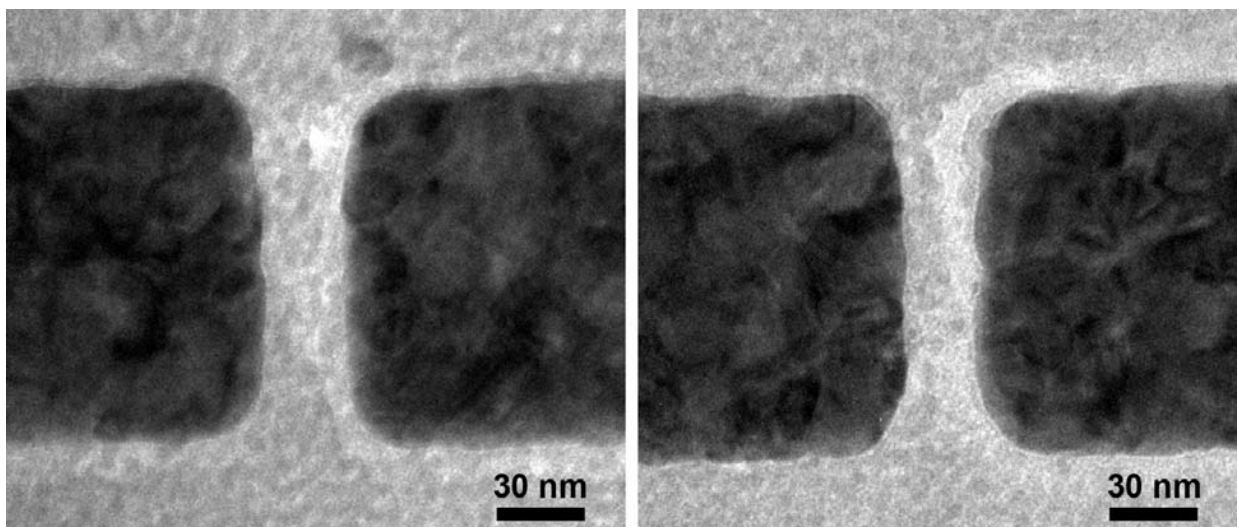

**Figure S10.** TEM images of two different nanogaps on the same chip after measurements with nanocrystals annealed at 498K.

## 8. Robustness of ratio *R* adjustment

For the initial I-V measurements of the devices taken using a 532 nm laser after several days without measurements, Figure S11 shows the measured photocurrent ratio *R* was in the approximate range of 1 to 10.

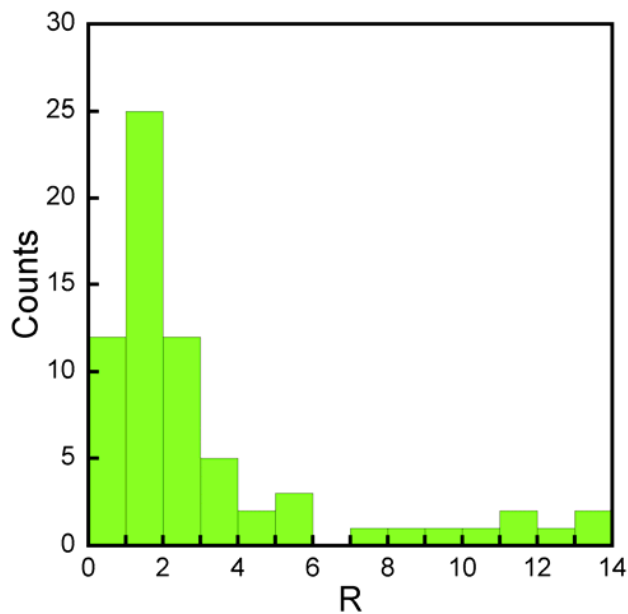

**Figure S11.** Distribution of *R* values for nanogap devices illuminated with 532 nm light. Histogram includes data from all treatments.

Illumination with a 650 nm laser produced photocurrent at room temperature (0.01-2.9 pA, with a mean of 0.36 pA) but lower photocurrent at low temperature (0.01-0.7 pA, with a mean of ~0.17 pA), yielding overall smaller ratios *R* in the range of 0.1 to 2. These two ratio populations are shown in a histogram in Figure S12. The smaller ratio *R* values for 650 nm versus 532 nm excitation can be understood by recalling the change in the absorption peak of these nanocrystals with temperature. At room temperature, the peak is thermally broadened which allows an overlap between the laser excitation and the absorption peak. The peak at 610 nm is thermally narrowed and blueshifted at low temperature because of the temperature dependence of the Stokes shift,[5] which reduces the overlap of the laser excitation and the peak, causing photocurrent to be lower. The photocurrent from illumination at 980 nm was also measured, but was found to be negligible as expected due to the energy mismatch between the nanocrystal bandgap and the energy of the incident photons.

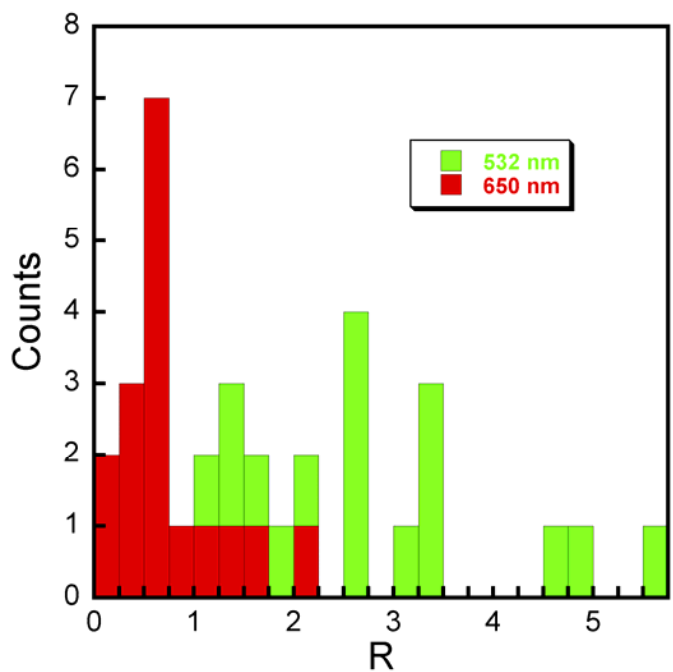

**Figure S12.** Histogram of the ratio $R = I_{78K}/I_{295K}$, with data taken from comparable treatments, for two laser excitation energies corresponding to excitation wavelengths of 532 nm and 650 nm. Higher illumination energy (532 nm) results in larger $R$, with an average $R$ of 2.8. The average $R$ for the 650 nm excitation is 0.7.

| Gap | *R* after no measurements | *R* after laser voltage treatment | *R* after dark voltage treatment | *R* after laser voltage treatment |
|-----|------|------|------|------|
| 5-1 | 0.38 | 1.81 | 0.72 | 0.94 |
| 5-2 | 2.67 | 4.44 | 1.20 | 1.43 |
| 5-3 | 0.36 | 1.00 | 0.33 | 0.48 |
| 5-4 | 1.63 | 2.85 | 0.53 | 0.75 |
| 5-5 | n/a | n/a | 0.75 | 1.25 |
| 5-6 | 1.5 | n/a | 1.5 | n/a |
| 5-7 | 1 | 0.88 | 0.31 | 0.59 |
| 5-8 | 0.75 | 1.56 | 0.43 | 0.72 |

| 5-9 | 0.69 | 1.88 | 0.23 | 0.75 |
| 5-10 | 0.85 | 3.29 | 0.51 | 1.27 |
| 5-11 | 0.095 | 1.06 | 0.04 | 0.20 |

**Table S1.** Relative photocurrent ratios of the low- and room-temperature photocurrents, $R = I_{78K}/I_{295K}$ for several nanogap devices on a single chip, illuminated with 650 nm light, increasing or decreasing with different treatments.

We find that using the laser voltage treatment to increase the relative photocurrent ratio $R = I_{78K}/I_{295K}$, or using the dark voltage treatment to decrease $R$, is a robust and repeatable process. Nanogap devices can be cycled to high and low $R$ values many times without the effect losing potency. We measured our devices for several months and continued to observe the same reversible behavior. Table S1 shows the relative photocurrent ratio $R$ of several nanogaps, and its changing value after laser or dark voltage treatments. This data was taken with a 650 nm laser.

## 9. Ratio $R$ inversion between $R < 1$ and $R > 1$

The temperature dependence of the photocurrent can be repeatedly reversed in *a single nanogap* device to yield a temperature-decreasing or temperature-increasing photocurrent. This is illustrated by one nanogap in Figure S13 for a range of laser intensities used.

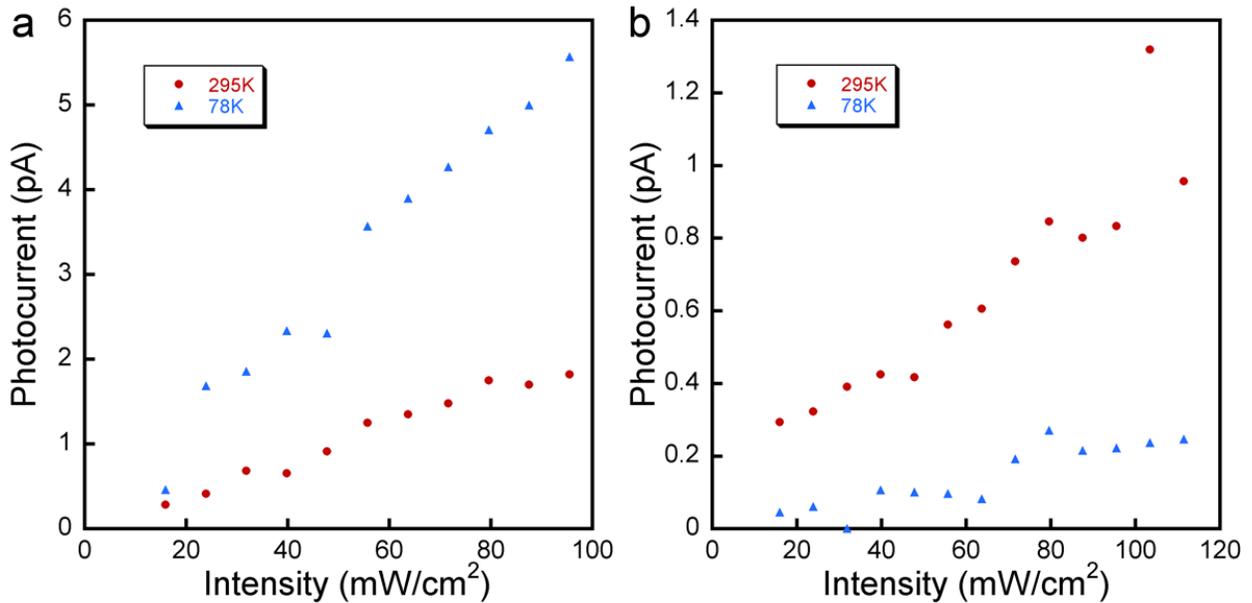

**Figure S13.** Measurements of photocurrent versus laser intensity (a) before and (b) after a dark voltage treatment of a single nanogap which show the ratio $R = I_{78K}/I_{295K}$ switching from (a) $R > 1$ when $I_{78K}$ (blue) > $I_{295K}$ (red) to (b) $R < 1$ when $I_{78K}$ (blue) < $I_{295K}$ (red) for all laser intensities used.